\documentclass[10pt,conference]{IEEEtran}
\IEEEoverridecommandlockouts
\usepackage{enumerate}
\usepackage[T1]{fontenc}
\usepackage{xparse}
\usepackage{enumitem}
\usepackage{tabularx}
\usepackage{scalerel}
\usepackage{cite}
\usepackage{amsmath,amssymb,amsfonts}
\usepackage{algorithmic}
\usepackage{graphicx}
\usepackage{textcomp}
\usepackage[dvipsnames]{xcolor}
\usepackage[font=small,labelfont=bf,tableposition=top]{caption}
\usepackage{dialogue}
\usepackage{caption}

\let\endorigquote\endquote
\renewenvironment{quote}{ 
  \vspace{-0.5\parskip}
     \leftmargin0cm   
     \rightmargin\leftmargin
}%
{\endorigquote}

\setlist[description]{
  font={\sffamily\bfseries},
  labelsep=0pt,
  labelwidth=\transcriptlen,
  leftmargin=\transcriptlen,
}

\newlength{\transcriptlen}

\NewDocumentCommand {\setspeaker} { mo } {%
  \IfNoValueTF{#2}
  {\expandafter\newcommand\csname#1\endcsname{\item[#1:]}}%
  {\expandafter\newcommand\csname#1\endcsname{\item[#2:]}}%
  \IfNoValueTF{#2}
  {\settowidth{\transcriptlen}{#1}}%
  {\settowidth{\transcriptlen}{#2}}%
}

\setspeaker{Streamer}
\setspeaker{Watcher}

\newcommand{\topanchor}[1]
  {\refstepcounter{#1}\addtocounter{#1}{-1}\par}
  
\def\BibTeX{{\rm B\kern-.05em{\sc i\kern-.025em b}\kern-.08em
    T\kern-.1667em\lower.7ex\hbox{E}\kern-.125emX}}
    
\begin{document}
%

\title{An Exploratory Study of Live-Streamed Programming
}
\author{\IEEEauthorblockN{Abdulaziz Alaboudi}
\IEEEauthorblockA{\textit{George Mason University} \\
Fairfax, Virginia, USA \\
aalaboud@gmu.edu}
\and
\IEEEauthorblockN{Thomas D. LaToza}
\IEEEauthorblockA{\textit{George Mason University} \\
Fairfax, Virginia, USA \\
tlatoza@gmu.edu}
}

%

\maketitle
\begin{figure}
\begin{minipage}{\textwidth}
  \includegraphics[width=\textwidth, keepaspectratio]{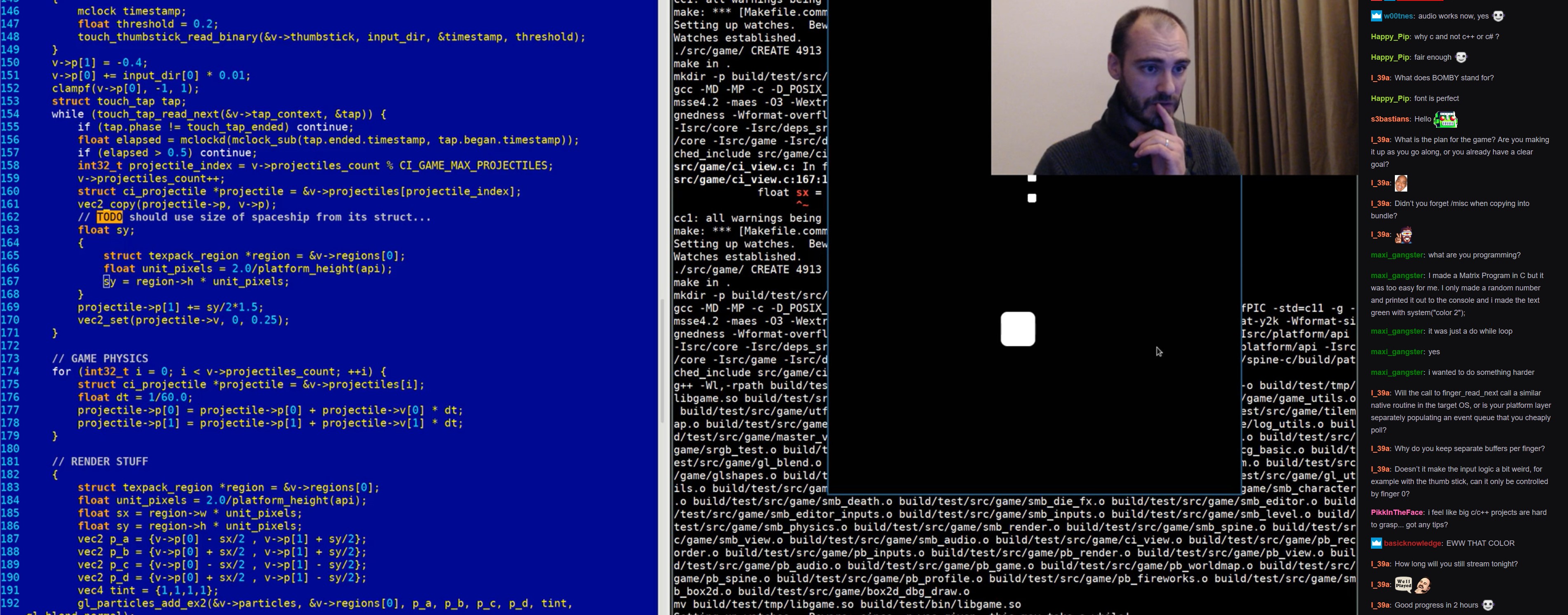}
  \captionof {figure}{In live-streamed programming, a developer working on a programming activity shares a live video of their work with other developers, who may interact through chat to ask questions, suggest alternatives, and help find defects.}
  \label{fig:teaser}
\end{minipage}
\end{figure}

%
\begin{abstract}
In live-streamed programming, developers broadcast their development work on open source projects using streaming media such as YouTube or Twitch. Sessions are first announced by a developer acting as the streamer, inviting other developers to join and interact as watchers using chat. To better understand the characteristics, motivations, and challenges in live-streamed programming, we analyzed 20 hours of live-streamed programming videos and surveyed 7 streamers about their experiences. The results reveal that live-streamed programming shares some of the characteristics and benefits of pair programming, but differs in the nature of the relationship between the streamer and watchers. We also found that streamers are motivated by knowledge sharing, socializing, and building an online identity, but face challenges with tool limitations and maintaining engagement with watchers. We discuss the implications of these findings, identify limitations with current tools, and propose design recommendations for new forms of tools to better supporting live-streamed programming.


\end{abstract}

%
%

%
%

%

\rule{0pt}{54.5ex} 

\section{Introduction}
Programming has long been in part a social activity, as developers share knowledge and collaborate on projects using platforms ranging from bulletin boards and mailing lists to GitHub, StackOverflow, and Twitter \cite{lakhani2004, Mamykina2011, singer2014Twitter,Dabbish2012,Vasilescu2013,Bosu2013, Tsay2014,MacLeod2018}. 
Recently, a new form of collaboration has emerged in which developers live-stream their work developing open source software \cite{Koebler2015}, inviting other developers to join a programming session and watch while engaged in activities such as writing code, debugging, and refactoring. In this paper, we refer to this form of collaboration as \textit{live-streamed programming}.
Live-streamed programming sessions are often first announced by a developer, sharing a date and time at which a session will begin using social media such as Reddit and Twitter. The announcement may also include a link to the session, hosted on a video streaming platform such as YouTube or Twitch. During the session, developers join, watch, and interact, using chat with the streaming developer to ask questions, offer suggestions, and warn about potential defects (Figure \ref{fig:teaser}).   


Live-streamed programming is increasingly common within open source, with thousands of developers streaming their work and tens of thousands watching \cite{Bernasconi2019}. Streamers use communities such as Reddit\footnote{https://www.reddit.com/r/WatchPeopleCode} and Shipstreams\footnote{https://shipstreams.com/} to connect with watchers. 
Developers write about their personal experiences with live-streamed programming, often comparing it to pair programming\cite{ Hinton2017, Bell2019, Stacoviak2018}. 


However, many questions remain about the characteristics of live-streamed programming and the motives of its participants. 
In this paper, we explore the nature of live-streamed programming as a new form of social coding and collaboration between developers.
Specifically, we focus on four research questions:


\begin{itemize}
    \item[\textbf{RQ1}] What are the characteristics of live-streamed programming?
    
 \item[\textbf{RQ2}]  To what extent is live-streamed programming a form of pair programming?
    
    \item[\textbf{RQ3}]  What motivates developers to host live-streamed programming sessions?

    \item[\textbf{RQ4}] What challenges do developers face in hosting live-streamed programming sessions?

\end{itemize}

To answer these questions, we conducted an exploratory study in which we watched 20 hours of live-streamed programming videos of seven developers (\textit{streamers}) working on a variety of software development activities and interacting with other developers (\textit{watchers}). We then sent a short survey to the seven streamers and received five responses with insight into their motivations and the challenges they faced. 

Our results suggest that live-streamed programming shares some characteristics and benefits with pair programming, but also differs in several key ways. Watchers interact with streamers less frequently and show less commitment. We identified two forms of challenges which developers face. First, streamers struggled to maintain engagement with watchers while working on development activities. 
Second, watchers were not able to easily explore the codebase and quickly onboard with the task streamers were working on, limiting their ability to collaborate effectively with streamers.
Based on these findings, we propose several design recommendations for how tools can more effectively support the live-streamed programming workflow.

\section{Background}

Social coding \cite{Dabbish2012, begel2010social} offers opportunities for developers to collaborate and share artifacts across organizations, teams, and individuals. Software companies and individual developers use platforms such as Github and GitLab to host open source projects and enable the creation of communities \cite{Lawrence2018, Asay2018}. Developers report that they enjoy the process of open sourcing their software and find it an effective way to improve their technical skills\cite{lakhani2004}. Researchers have studied how developers discuss and evaluate code contributions on platforms such as Github \cite{Tsay2014}, identifying challenges and best practices regarding code review \cite{MacLeod2018}. Developers use communities such as StackOverflow to share knowledge and help each other by answering programming-related questions \cite{Mamykina2011}. Social coding also occurs through \textit{Screen casting}. Screencasting differs from live-streamed programming in that developers work solo and then post their video for asynchronous consumption. It is often used to document and share knowledge via online video sharing platforms such as YouTube, helping developers to promote themselves by building an online identity \cite{macleod2015}.

Pair programming is a social and collaborative activity in which two developers sit together at one computer, collaborating on designing, coding, debugging, and testing software. The collaboration between the two developers occurs by each taking two roles interchangeably. The first role is the \textit{driver}, whose responsibility is to act, including writing code, debugging, and testing. The \textit{navigator} is responsible for observing the driver and suggesting strategies, alternative solutions, and possible defects\cite{williams200Book}. Several studies have investigated the effectiveness of pair programming in both industrial and educational contexts \cite{Hulkko2005}. Pair programming has been found to enable students to produce better programs \cite{McDowell2002} and to achieve better grades \cite{Nagappan2003}. In industrial settings, studies suggest that pair programming provides benefits over traditional solo programming. Professional developers report creating higher quality code, inserting fewer defects \cite{Cunningham2000}, and an improvement in team communication \cite{Williams2000} when practicing pair programming. One form of pair programming is mob programming \cite{Dalton2019}, in which a whole development team works at a single computer. The role of the driver is rotated among team members every 5-15 minutes. This practice has been found to increase team productivity and encourage face to face communication \cite{zuill2016mob}. 


Distributed pair programming is a form of pair programming in which the driver and the navigator are not collocated and work remotely\cite{baheti2002}. For distributed pair programming to have the same benefits as collocated pair programming, tools that support ``cross-workspace information infrastructure'' must exist \cite{Flor2006} to ensure that the driver and navigator communicate effectively \cite{Canfora2006}. Several tools have been proposed to support distributed pair programming \cite{da2015distributedSurvey}. For example, Saros\cite{Salinger201} and XPairtise\cite{tsompanoudi2013} are Eclipse plug-ins that help developers to communicate during the pair programming session. Saros is a flexible tool for distributed pair programming, providing a synchronized workspace and visualizations to enhance awareness between the driver and the navigator. XPairtise is built to support students' collaboration by offering classroom support capabilities to add programming and group assignments. 

Live coding is a social coding practice in which the goal is to perform and demonstrate in front of an audience, often constructing small programs with 20 to 50 lines of code \cite{blackwell_et_al-Live-Coding}.
Live programming often occurs to produce electronic music in front of an audience \cite{nilson2007live, collins2003live, magnusson2011algorithms} and to teach programming to students during lectures \cite{Barker2005, ChenLAS2019}. 
Live-streamed programming differs from live coding in that it shows developers engaged in software development activities on projects that are intended to be production quality software (e.g., Shipstreams). 




\begin{table*}
\caption{A summary of the ten sessions. The projects' number of contributors and lines of code (LOC) indicate the size and level of activity of each project. Each project URL is prefixed with https://github.com/.}
\label{tab:table1}

 \begin{tabular*}{\linewidth}{c|c|c|c|c|c|c} 
 \hline
 Session&Streamer&Main Software Development Activity& Contributors&LOC& Project URL &Duration Obs. \\ 
 \hline\hline
  1&D1 & Implementing Features& 5 & 385 Python & eleweek/SearchingReddit& 2:57:35 \\ 
 \hline
   2&D1 & Implementing Features & 5 & 385 Python& eleweek/SearchingReddit  & 1:49:00  \\
 \hline
  3&D2 & Implementing Features& 1 & 423 C++ & benhoff/face-recognizer-gui  &1:25:28  \\
 \hline
   4&D2 & Implementing Features & 1 &  423 C++ &benhoff/face-recognizer-gui & 2:02:36  \\
 \hline
  5&D3 & Code Refactoring& 41 & 32.6K Rust & graphql-rust/juniper &2:07:00  \\
 \hline
  6&D4 & Responding to Issues and Pull Requests & 299 & 90.3K JavaScript & pouchdb/pouchdb & 2:37:00  \\ 
 \hline
 7&D5 & Implementing Features & Unknown & Unknown C & Unknown & 2:26:30  \\ 
 \hline
  8&D6 & Debugging & 37 & 8.4K JavaScript& wulkano/kap& 1:21:06   \\
 \hline
   9&D6 & Debugging & 50 & 8K JavaScript& buttercup/buttercup-desktop& 1:31:24  \\  
 \hline
 10&D7 & Code Refactoring & 24 & 14K JavaScript & noopkat/avrgirl-arduino & 1:29:37  \\
 \hline

\end{tabular*}
\end{table*}

Prior studies of live-streamed programming have explored its educational implications. Haaranes \cite{Haaranen2017} observed one streamer during three sessions building a game and proposed teaching methods that expose computer science students to larger software projects. Faas et al.  extended this work with two studies \cite{Faas2018, Faas2019} involving observations and interviews of streamers on Twitch. This work revealed the mentoring relations that occur in live-streamed programming and its potential for use as a learning platform. In this paper, we build upon this work by exploring live-streamed programming as a form of social coding and collaboration. In particular, we investigate its characteristics, how it differs from pair programming, and the motivations and challenges of its participants. 


\section{METHOD}
To answer our research questions, we first selected 20 hours of live-streamed programming sessions. To identify these sessions, one approach would be to wait for streamers to announce live-streamed programming sessions through social platforms and then join to observe in real time. However, this limits observations to new sessions. As we did not wish to actively participate, this offered no benefits. We thus chose to search for archived sessions on platforms such as YouTube, Twitch, and Reddit. We used two inclusion criterion to select sessions. First, each session should show a streamer engaged in a software development activity, such as writing, debugging, or refactoring code. Second, each session should show a streamer and watchers interacting via voice or text, and these interactions should be accessible via the video or chat history. We searched for sessions that satisfied these criteria and selected ten, all of which were hosted on YouTube. 
We did not include videos hosted on Twitch and Reddit, as these videos may not be archived. For examples, Twitch only archives videos for 14 days \footnote{https://help.twitch.tv/s/article/videos-on-demand,}.

The 20 hours we selected included ten videos by 7 streamers (six male, one female; D1-D7) working on both personal open source projects (D1, D2, D5, D7) as well as contributing to popular open source projects (D3, D4, D6). Table \ref{tab:table1} summarizes the sessions. We watched each of these sessions, identifying common activities among the streamers and watchers. We then compared our observations with descriptions of pair programming reported in prior work \cite{williams200Book}. 

To further understand the challenges and motivations of streamers, we sent a brief two question survey to the seven streamers, focusing on their motivations and challenges. 
Five responded (D1-D5).  The sessions links and the data we collected from the sessions are publicly available\footnote{https://github.com/Alaboudi1/live-streamed-programming}.


\section{RESULTS}
\subsection{What are the characteristics of live-streamed programming?}

Our observations of the ten live-streamed programming sessions revealed common characteristics in how developers \textit{advertise}, \textit{start}, \textit{plan}, \textit{use}, and \textit{end} sessions. Streamers first \textit{advertised} their sessions through announcements on Twitter (D3, D4, D6, D7) or Reddit (D1, D2, D5), including the expected time of the streaming session and a link to the session. All except session 9 had a posted announcement before the start of the stream. 
 
Each session \textit{started} with developers showing their face and giving an introduction and background for the session. This lasted an average of 9 minutes (range 1-18 minutes).  Streamers educated watchers about the codebase, technologies used, and the purpose of the session. All sessions  were streamed from a private space except session 10, which occurred in a public coffee shop.  

Streamers often began their work by offering a general \textit{plan} of what they would do. D5 stated that he would build a game but he ``\textit{do[es] not know what the game exactly will be}''. D4 streamed the process of debugging issues reported in open source projects he maintained. Not all the issues were resolved during the session, as he stopped while debugging some and moved on to others. This left the live-streamed session without a clear goal, leading some watchers to ask when the session would end.

After finishing the introduction, the streamers \textit{used} the session to engage in development activities, including implementing features, debugging, refactoring, and searching for documentation and StackOverflow posts. Watchers interacted with streamers during these activities by asking questions and helping the streamer with the task at hand, which sometimes prompted new behavior by the streamers. For example, when watchers posted a library to use or a link to read in the chat, streamers almost always stopped their current activity to check what the watchers suggested. Streamers (D1, D4, D6) were interrupted by people in their physical environment who brought drinks, talked with them, or rang their doorbell. While most streamers waited until the end of their session to commit their code changes to their version control system, D1, D3, and D4 continually pushed their changes throughout the sessions, enabling watchers to run and test them.

While streamers were engaged in development activities, watchers continually joined and left throughout. When they first joined, watchers sometimes communicated with the streamer or with other watchers through chat. However, most watchers remained passive participants. For example, session 1 had over 50 watchers but only 12 who sent at least one chat message.

Streamers often \textit{ended} the live-streamed programming session by giving a demo of the work they had completed. They also pushed their final code changes to a public repository for the watchers to view. Finally they often concluded by stating the tentative time and date, if any, of their next session.

\subsection{To what extent is live-streamed programming a form of pair programming?} 

\begin{figure}
    \includegraphics[ width=\linewidth, keepaspectratio, clip]{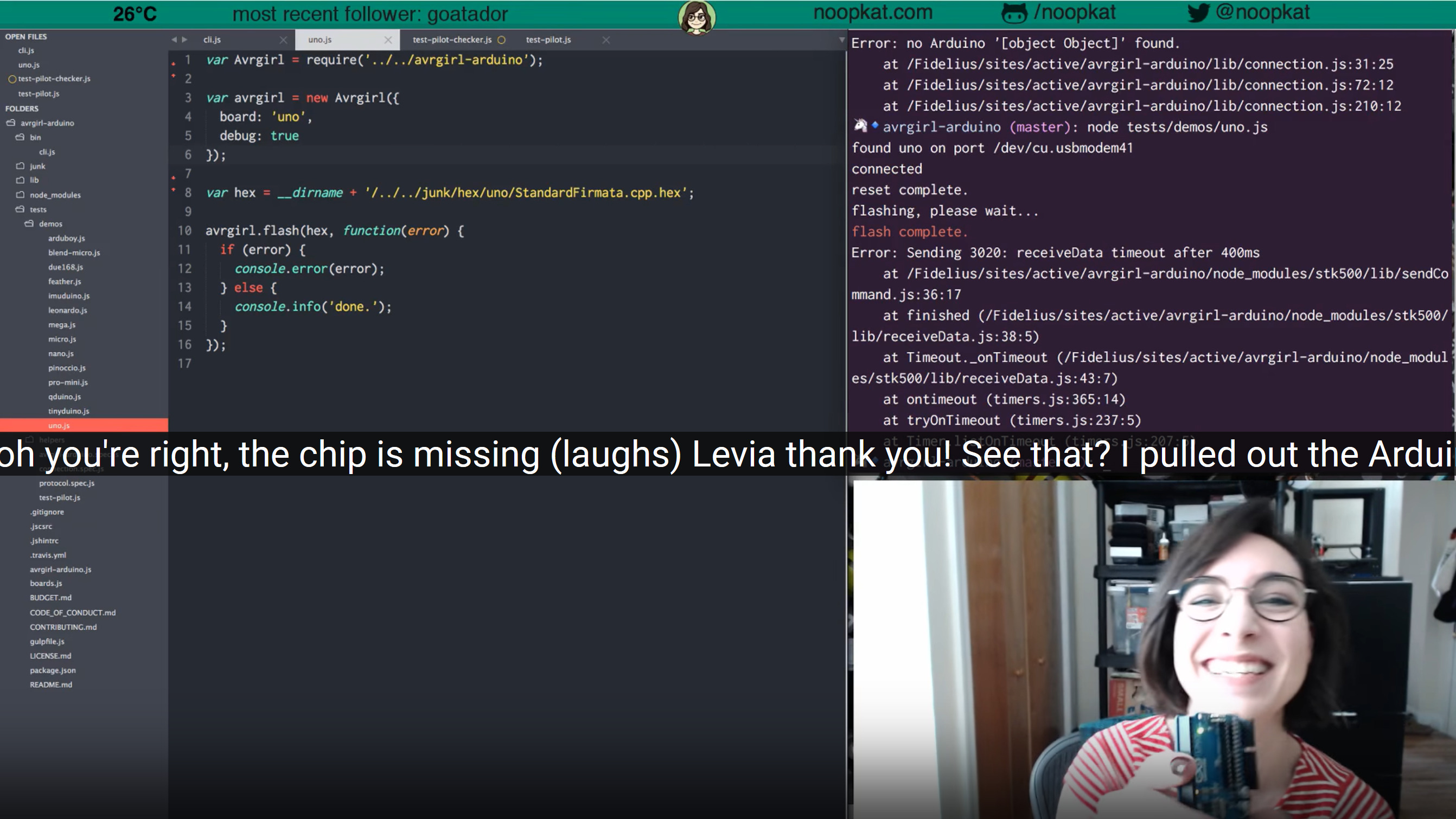}
    \caption{ \label{fig:example2} A streamer (D7) thanked a watcher for finding a defect.}
    
\end{figure}

\begin{figure}
    \includegraphics[width=\columnwidth, keepaspectratio, clip]{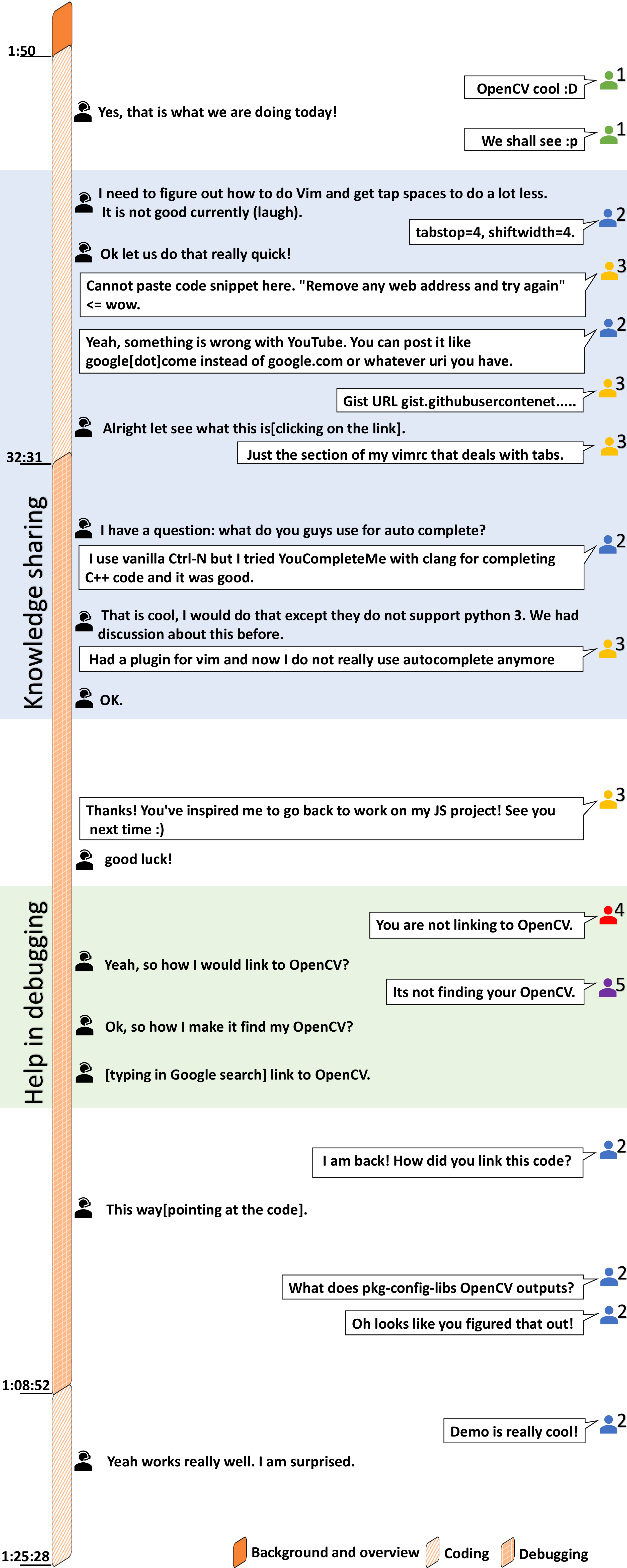}
    \centering
    \caption{\label{fig:studyExample} Excerpts from session 3 which offer examples of the interactions between the streamer (left) and watchers (right). These excerpts illustrate knowledge sharing between the streamer and the watchers, watchers helping in debugging, and watchers joining and leaving throughout the session. }
\end{figure}

In \textbf{pair programming}, there are two distinct roles. The role of the driver is to take action, while the navigator supports and guides the driver \cite{williams200Book}. In the live-streamed programming sessions we observed, the behavior of the streamers and watchers generally followed these roles. Streamers wrote code, debugged, and designed, while watchers observed the streamer, helping debug, suggesting alternative solutions, and offering tips. 
However, effective pair programming requires the driver and navigator to interact at least once every 45 to 60 seconds \cite{williams200Book}. In our observations, we did not find that the streamers and watchers collaborated at this level of frequency. This made live-streamed programming different from pair programming in two key ways. First, streamers did not actively solicit input from watchers. Throughout the sessions, streamers educated watchers about the project and explained design decisions and technology choices they made. When watchers helped streamers debug or find better tools and solutions, streamers thanked the watchers, as they were not expecting such a contribution (Figure \ref{fig:example2} ). Second, watchers exhibited weaker focus and commitment to the task at hand.  Watchers joined and left throughout the sessions, inspecting and working on other parts of the code if it was available to them, and interrupted the streamer by asking questions that were not relevant to the current task or which had been answered before. Figure \ref{fig:studyExample} lists examples of interactions between the streamer and watchers from session 3.

Studies have found that pair programming offers several benefits, such as producing high-quality code, shortening development time, facilitating knowledge sharing, and creating joy in the work environment \cite{williams200Book}. The observations and survey responses suggest that live-streamed programming may offer some of these same benefits.
We were able to identify seven instances in which watchers improved the quality of their code, reporting three defects and helping debug another four. D1 stated during session 2 that he would not have discovered a defect a watcher reported, as it was related to an edge case that he was not aware of: ``\textit{Thank you! I would not have thought about the accented character}''. Another streamer (D5) tested multiple incorrect hypotheses about the cause of a defect before a watcher suggested a correct hypothesis.

 

\begin{dialogue}
    \speak{\textbf{Streamer}} [After reading the error] I think I know why, but before I say it I want to confirm it.
    \speak{\textbf{Watcher}} Did you forget /misc when copying into a bundle?
    \speak{\textbf{Streamer}} [Stopped debugging and started reading the chat] I do not think so, but thanks! Let's see.
    \speak{\textbf{Streamer}} [After testing the watcher's suggested hypothesis] Bravo, you are right! I forgot to do that, thank you!
\end{dialogue}

We observed 38 instances in which streamers and watchers shared tips, explaining technical topics and offering alternative technology choices to each other. D4 was explaining his git workflow to the watchers, and one of the watchers pointed out a better workflow that he did not know. ``\textit{[Reading the tip message from the chat] Oh thank you, Jake! Let us try it out. [He tries it.] OK, that is a lot faster than [the] three steps that I was doing before, thank you! That is going to improve my efficiency a lot}''. D3 was working on refactoring an existing codebase in an open source library, where the maintainer of the library was among the watchers. Both the streamer and the watcher (maintainer) asked and answered each others' questions while the streamer was working. When the streamer finished working on the project, the watcher wrote ``\textit{it has been very useful, learned new things, thanks!}''.  

One advantage of pair programming that we did not observe was shortening development time.  Two developers (D2, D4) explicitly stated in their sessions that they usually take less time doing the same job offline. ``\textit{I really [am] only keeping doing this because people say, hey, I like the thing you are doing. Otherwise, I really would not do it, because I would code faster without [it]}'' - D2. One contributing factor might be a lack of effective tooling, which we discuss in RQ4 below.

\textbf{Mob programming} is a form of pair programming in which more than two developers collaborate and rotate the role of driver between them. We did not observe any instances of a streamer and watchers rotating their roles.

\subsection{What motivates developers to host live-streamed programming sessions?}

From our observations of the sessions as well as the survey responses, we identified three motives for live-streamed programming: \textit{knowledge sharing}, \textit{socializing and work enjoyment}, and \textit{building an online identity}.  

We found \textbf{knowledge sharing} to be a common activity in live-streamed programming, confirming findings from prior studies \cite{Faas2018, Faas2019, Haaranen2017}. We identified several instances in which streamers and watchers explained to each other design recommendations and choices, alternative tools and libraries, and tips they found to increase their productivity. D3 stated at the beginning of a live-streamed programming session that he does the recording for people who `` \textit{want to learn or see the language [Rust] being used in a more advanced way}''. He further expanded on this motivation in his survey response:

\begin{quote}
 ``\textit{I find it very rewarding to hear that others are learning from 
    what I'm putting out there. I've always enjoyed teaching, and 
    this medium seems to work very well for a lot of people.}''
\end{quote}

D4 reported in his survey response that what motivated him is ``\textit{to show``a daily in the life''of an open-source maintainer}''. He pointed us to a blog post that he wrote about the struggles open source software maintainers of popular library experience day to day \cite{Lawson2017}. During his session, he went through 69 separate Github issues and pull requests opened by the community, showing how he answers these issues and reviews code written by other developers.

D1 indicated that, besides his live-streamed programming sessions, he also joins other developers' sessions, collaborating with them and assisting novices while developing software.

\begin{quote}
 ``\textit{My favorite aspect of streaming was collaboration. I'd even say that more than my own streaming I enjoyed helping other people on their streams more - especially novices.}''- D1
\end{quote}


\noindent\textbf{Socializing and work enjoyment} was another key motive. D5 stated that what made him start to host live-streamed programming sessions was because he ``\textit{enjoyed watch[ing] others coding [...] and thought it would be fun as an experiment to try to do that too}''. D3 stated that socializing during live-streamed programming session made him enjoy his development work to a degree that he would work for 5-6 hours continuously, and that he often uses it to motivate him to work on his research. 

\begin{quote}
``\textit{I genuinely enjoy the experience of programming with other 
    people. It feels like we're solving something together, and having them there makes the entire experience more fun. [...] That's amazing! 
    Often I will choose projects that tie into my research too, so 
    it's even useful work.}'' - D3
\end{quote}

Watchers expressed enjoyment in the process of sharing, watching the development workflow, and collaborating with streamers. One watcher expressed enjoyment after suggesting an idea to the streamer during session 1. ``\textit{This is so cool... :) watching others code and contributing with ideas}''. Another watcher enjoyed the idea of watching another developer while coding. ``\textit{It is really like hardcore watching [another] hardcore coding like this. I love it!}'' (session 2).

\noindent A third motive was \textbf{building an online identity}. D1 stated that a recruiter from Google had contacted him because of his streaming activity. D2 has a similar experience where he ``\textit{had multiple new job opportunities that came out of streaming online}''. D3 expressed no interest in job opportunities, as he was still in graduate school, but stated that ``\textit{with the videos I think there's now a lot more people who know who I am}''. D4 pointed out that although his online identity had not grown because of his streaming activity, he was happy that he inspired another developer to become a streamer (D7), who later landed a job at Microsoft because of her streaming activity \cite{Hinton2017}. 

\subsection{What challenges do developers face hosting live-streamed programming sessions?}

Our observations and survey responses revealed two key challenges that streamers and watchers experienced. 

\noindent\textbf{Tools limitations}. Our observations suggest that watchers sometimes help streamers while debugging, and thus may need access to the source code to inspect and test their hypotheses. These needs were not directly supported by any of the development tools used by the streamers. For example, a watcher was trying to help D7 while debugging, but the interaction was slow and consisted of ``\textit{Are you referring to this line?}''. This largely was due to a lack of direct access to the code. To answer a question, he needed first to write the question down in chat and wait for the streamer to read it and respond, which sometimes then required a request for clarification. D7 referred to this interaction as ``\textit{slow, delayed pair programming}'' while the watcher felt guilty for not being able to help: ``\textit{oh my gosh I am terrible}''.

D1 stated in the survey that the lack of tools which ``\textit{remove some friction with figuring out what the streamer is doing right now and allow more collaboration}'' created a challenge in maintaining productive live-streamed programming sessions with watchers.

\begin{quote}``\textit{I really wish there was a way of letting viewers browse your code independently, and maybe even a way of quickly recreating the same dev environment in some way.}'' - D1
\end{quote}
    
In addition to giving watchers the ability to explore the code independently, watchers may need help in understanding the codebase and the current task of the streamer, particularly when they join a session late. D2 reported that existing tools lack support for giving a quick overview of the project and its architecture for watchers. 

\begin{quote}
 ``\textit{I think the hardest thing the viewers would have was understanding the architecture of the code base, or how things fit together. So if you can figure out how to keep track of the problem that you are working on, visually represent the code base/changes made against the task, and how that code fits together, you would have a good visual editor.}'' - D2
\end{quote}


\noindent\textbf{Maintaining engagement}. Writing and debugging code are cognitively demanding tasks which may often require the developer's full attention. Streamers often stayed silent while they were thinking about what might cause a defect in their code. This may create a disconnect between the streamer's and watchers' mental models, leaving watchers less engaged in the current task.

\begin{quote}
 ``\textit{I think the hardest thing is to make sure you keep the 
    viewers engaged. Don't just code without speaking -- you need to voice your thoughts, and ask the people watching the same questions you'd normally mull over in your head while programming. Otherwise 
    they won't learn anything as it'll be an entirely passive 
    experience. And I don't think people are generally used to speaking 
    their thoughts out loud while programming!}'' - D3
\end{quote}

Another streamer reported that coding while a camera was recording impacted his ability to maintain watchers' engagement.

\begin{quote}
 ``\textit{While coding, there are many other thoughts that need to be managed. You literally have a camera on you while you are coding, and that can greatly affect my focus. Likewise, once I am concentrating on some code - I may forget about the camera and not explaining my thought-process clearly}'' - D5
\end{quote}

\section{Limitations and threats to validity}
Our study has several important limitations and potential threats to validity. One potential threat to external validity is how representative the ten live-streamed programming sessions that we selected were. We included sessions based on the development activities and the presence of interactions between  streamers and watchers. However, our selected sessions were also diverse in programming languages, projects size (ranging from 385 to 90.3K lines of code), and the number of developers contributing to the projects. Another potential threat to the generalizability of our results is the size of our sample. In our study, we observed a total of ten sessions which spanned 20 hours of activity by 7 streamers.
Prior work \cite{Robillard2004, Abi-Antoun2010, Sillito2008, Starke2009} which involved analysis of developers working in various development activities had on average 6.6 developers (range 3-10) and an average of 8 hours of videos (range 5-15). 

A potential threat to internal validity is related to our qualitative data analysis. As our focus was to identify the nature of live-streamed programming work rather than characterize the frequency of specific practices or characteristics, we employed a lightweight data analysis and did not code most events for their frequency.


\section{DISCUSSION}
Live-streamed programming is a new form of collaboration in which streamers invite watchers to observe them perform programming work. Our observations revealed that developers' roles, interactions, and benefits have similarities to pair programming. However, live-streamed programming differs in two important ways.
Streamers and watchers interactions were less frequent, and watchers were less committed to the task at hand, joining and leaving throughout the session. Streamers were motivated by the opportunity to use their work as an educational resource and as a means to promote themselves, confirming findings in prior studies \cite{Faas2018, Faas2019, Haaranen2017, macleod2015}. One additional motive we observed was that both streamers and watchers enjoyed the process of streaming and watching coding activities. D3 reported that streaming helped him to enjoy working for hours for different projects, including his academic research work. Further research on this motive may help to better reveal how live-streamed programming might be used create an enjoyable and stimulating working environment, assisting developers in enjoying daunting coding activities and potentially increasing their productivity.


One unique characteristic of live-streamed programming is that, unlike the navigator in pair programming, watchers may or may not contribute to the task at hand and may join and leave throughout the session. Current tools that support distributed pair programming such as Saros\cite{Salinger201}, XPairtise\cite{tsompanoudi2013}, and Visual Studio Live Share\cite{Silver2017} were not designed to support this form of work. For example, they do not offer support for watchers who join late in gaining a holistic view of what the streamer is working on and planning for the rest of the session. Further, existing pair programming tools require investment and commitment by the watcher to install and configure the development environment to run them. Online development environments such as Codenvy\footnote{https://codenvy.com}, Codeanywhere\footnote{https://www.codeanywhere.com}, and Cloud9\footnote{https://aws.amazon.com/cloud9} may require less setup. But they may offer development tools and workflow that differ from what the streamers use locally. They also lack other features to support the workflow of live-streamed programming, discussed below.

The lack of effective tooling for live-streamed programming has already motivated some open source developers to invent new tools that address some of the challenges we identified. During the time in which this study was conducted, an open source extension for VS Code\footnote{https://github.com/clarkio/vscode-twitch-highlighter} was introduced which allows watchers to refer to a location in the source code via chat, which results in an in-editor visualization for the streamer. While addressing one key aspect of streamer and watcher collaboration for one specific development environment, there remain additional challenges to be addressed such as how to enable watchers to explore the code independently, how to help them with onboarding with the streamer's development work, and how to help streamers maintain watchers' engagement. We discuss design implications for tools in the following section.
\begin{figure*}
    \centering
    \includegraphics[ width=150mm, keepaspectratio, clip]{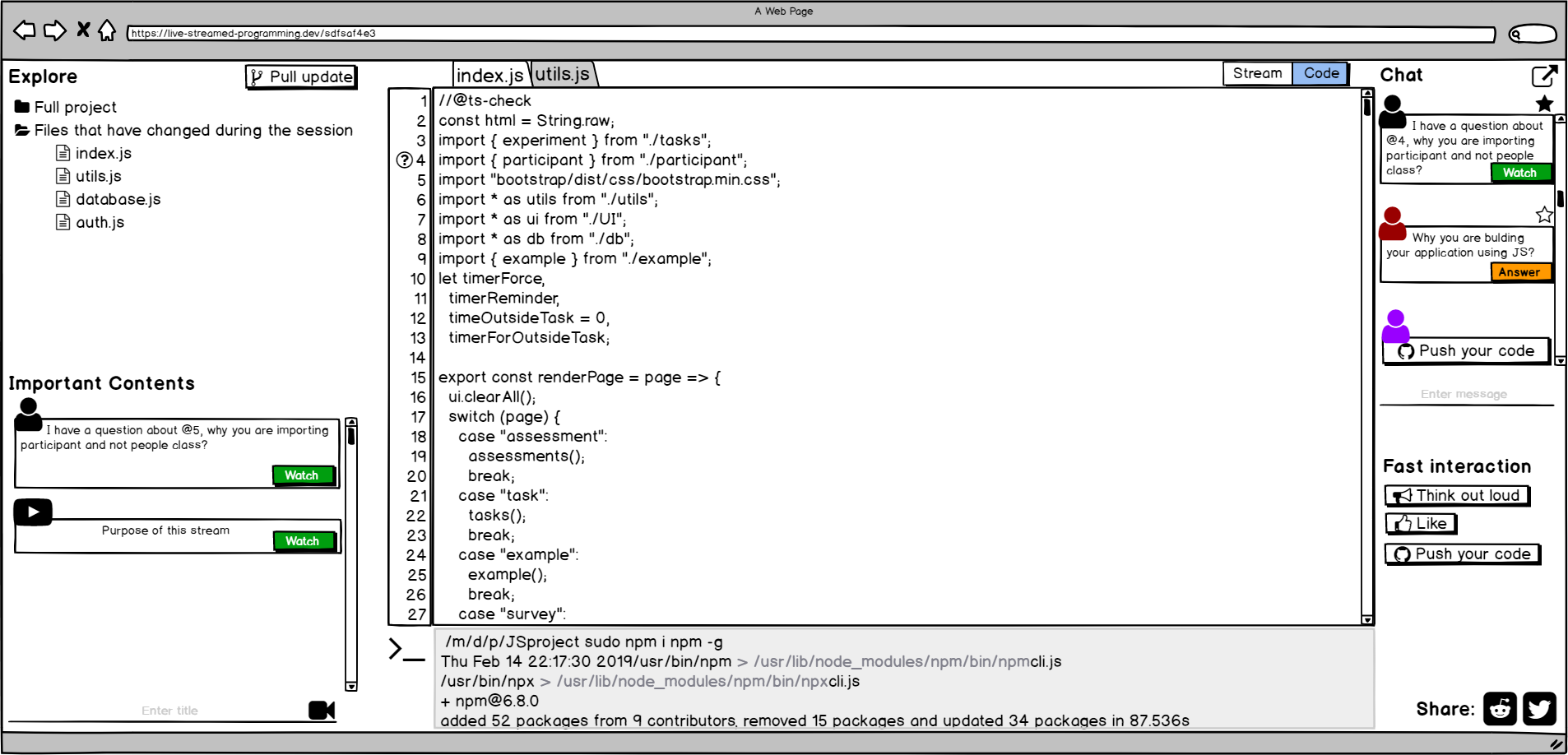}
    \caption{A tool mockup illustrating our design recommendations}
    \label{fig:mockup}
\end{figure*}
\section{Design Implications}
Our study revealed that streamers sometimes struggle in \textit{maintaining engagement} with watchers, while watchers need tools that help them \textit{explore code} independently and help \textit{onboard} to the streamer's project. In this section, we propose a set of design recommendations for addressing these needs.

\subsection{Maintaining Engagement}
Instead of relying exclusively on the streamer to continually remember to keep watchers engaged while immersed in development activities, watchers might instead help in quickly notifying the streamer of their needs. For example, watchers may want to inform the streamer to think aloud to make them follow along and avoid making them speculate as to why she acted as she did. We call these notifications from watchers \textit{Fast interactions}.

\textit{Fast interactions} are a way for watchers to communicate common needs to the streamer quickly. Instead of writing a text message to remind the streamer to think aloud, watchers might click a button to notify the streamer to talk through what she is currently doing. Watchers may need access to the latest version of the code or to increase the font size to make the code in the streamer's editor more easily readable. Watchers may also want to give a sign of support to inform the streamers that they are following along. Faster interactions might be triggered by either the streamer or the watcher. This form of communication may also encourage more watchers to actively collaborate by lowering the barriers to participate.



\subsection{On-demand Code Exploration}
Our findings suggest that watchers may benefit from being able to explore the code independently from the streamers.
However, the latest version of the codebase may not always be accessible for the watchers or may require the installation of several dependencies and modification to the environment variables to build the project and run it. Such barriers increase the cost of code exploration for the watchers and may decrease their engagement with the streamer. 
This suggests that a tool to support live-streamed programming should offer on-demand code exploration for watchers. Additionally it should not require any tool or development environment set-up for either the streamer or for the watchers.

One approach may be to build a web application that connects to the streamer's project repository (e.g., Github) and offers the necessary environment for the watchers to build and run the software. 
Watchers may also need to know what files have been changed recently so that they can only focus on these files instead of the entire project.

\subsection{Fast Onboarding}
One difference between pair programming and live-streamed programming is that, unlike navigators, watchers join and leave throughout the session. Watchers who join later may have questions that have already been answered about the purpose of the project, previous design decisions, and the plan for the rest of the session. To help watchers quickly answer these questions and get up to speed, tools might offer \textit{content linking} that help watchers quickly traverse the history of the session to find answers for their questions.

Tools such as chat.codes offer the ability to link between code fragments and natural language descriptions in ``asynchronous settings where one user writes an explanation for another user to read and understand later'' \cite{Oney2018}. Our proposed feature for content linking extends this idea to consider linkages between video segments, code fragments, and natural language text in the chat. Streamers might indicate video segments in the stream that are of importance to watchers who came late. These video segments may include the introduction to the session or answers to key questions posed in the chat. Besides video segment linking, watchers who have questions about the codebase should have the ability to link their questions with a code location, helping offer both the streamer and other watchers more context.

\subsection{Tool Workflow}
To illustrate how our design recommendations might be embodied in a tool for supporting live-streamed programming, Figure \ref{fig:mockup} depicts a mockup of a web-based tool for live-streamed programming.  
Before the live-streaming session begins, the streamer first logs into the tool website and provide links to both the live-streamed programming session and the public repository that hosts the source code. The tool then aggregates access to the session and the code repository in one web page. The streamer may then advertise the session by using the share button or copying the page link and posting it to social platforms. After the session begins, the streamer may work in her preferred development environment and push any updates to the code repository for the watchers to use.

When watchers join the session, they have the option to either watch the live stream or independently explore the codebase, triggered through a toggle button (\scalerel*{\includegraphics{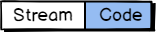}}{B}). Invoking the code mode brings up a lightweight development environment containing a code editor, terminal, and file browser. In the file browser, watchers can pull updates from the repository, explore the entire project, or limit the focus to the files that have been recently updated during the session. Watchers may interact with the streamer via chat or fast interaction (bottom right in Figure \ref{fig:mockup}). In the regular chat, watchers can link to a code location in the chat using \textbf{@}. This creates a (\scalerel*{\includegraphics{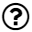}}{B}) mark on the gutter at that location so that other watchers can see the presence of discussions by skimming the gutter. 

When answering a question posted in the chat, the streamer can mark the beginning of the answer by clicking on (\scalerel*{\includegraphics{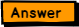}}{B}), linking the question to its answer to make it easy for latecomers to find the answer by clicking on (\scalerel*{\includegraphics{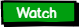}}{B}). Both the streamer and watchers may mark important chat questions by clicking on the (\scalerel*{\includegraphics{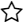}}{B}) and indicate segments of the session where the streamer discusses important topics by clicking on (\scalerel*{\includegraphics{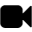}}{B}). This might be used to create a list of important questions and moments for late watchers to explore. 






\section{Conclusion}
Live-streamed programming is a form of collaboration in which a streamer synchronously broadcasts their programming work and interacts with watchers. Our results offer insight into its characteristics, why developers use it, and the challenges they face. Our findings suggest that live-streamed programming shares several characteristics and benefits with pair programming, but also differs in ways which make existing distributed pair programming tools less effective. We propose several design recommendations for how future tools can offer more effective support. We believe that such tools may enable more developers to make portions of their development work public and accessible for other developers to observe and collaborate.

%
\section*{Acknowledgments}
We would like to thank the survey respondents for their time. This research was funded in part by NSF grant CCF-1703734. The first author is supported in part by a King Saudi University Graduate
Fellowship.


\bibliographystyle{IEEEtran}
\bibliography{main}
%

%

\end{document}